\documentclass[10pt,conference]{IEEEtran}

\usepackage{amsmath}
\usepackage{amssymb}

\newtheorem{Thm}{Theorem}
\newtheorem{Lem}{Lemma}

\begin{document}

\IEEEoverridecommandlockouts

\title{Further Results on Coding for Reliable Communication over Packet
Networks
\thanks{This work was supported by the National Science
Foundation
under grant nos.\  CCR-0093349, CCR-0325496, and CCR-0325673, by
Hewlett-Packard under grant no.\  008542-008, and by the Air Force
Office of Scientific Research under grant no.\  F49620-01-1-0365.}}

\author{\authorblockN{Desmond S. Lun and Muriel M\'edard}
\authorblockA{Laboratory for Information and Decision Systems\\
Massachusetts Institute of Technology\\
Cambridge, MA 02139, USA\\
E-mail: \{dslun, medard\}@mit.edu}
\and
\authorblockN{Ralf Koetter}
\authorblockA{Coordinated Science Laboratory\\
University of Illinois\\
Urbana, IL 61801, USA\\
E-mail: koetter@uiuc.edu} \and
\authorblockN{Michelle Effros}
\authorblockA{Dept.\  of Electrical Engineering\\
California Institute of Technology\\
Pasadena, CA 91125, USA\\
E-mail: effros@caltech.edu}
 }

\maketitle


\begin{abstract}
In ``On Coding for Reliable Communication over Packet
Networks'' (Lun, M\'edard, and Effros, \emph{Proc.\  42nd Annu.\
Allerton Conf.\  Communication, Control, and Computing}, 2004), a
capacity-achieving coding scheme for unicast or multicast over lossy
wireline or wireless packet networks is presented.  
We extend that paper's results in two ways:
First, we extend the network model to allow 
packets received on a link 
to arrive according to any process with an average
rate, as opposed to the assumption of Poisson traffic with i.i.d.\  losses
that was previously made.
Second, in the case of Poisson traffic with i.i.d.\  losses, 
we derive error exponents that
quantify the rate at which the probability of error decays with coding
delay.
\end{abstract}

\section{Introduction}

This paper extends \cite{lme04}, which describes a coding scheme
for reliable communication over packet networks.  The most notable
features of this scheme are that it is capacity-achieving; that 
intermediate nodes perform additional coding yet 
do not decode nor even wait for a block of packets before sending out
coded packets;
that little explicit network management is required, with little or no
requirement for feedback or for co-ordination;
and that all coding and decoding operations have polynomial complexity.
The scheme therefore differs from 
schemes based on the digital fountain approach
(e.g.,\  \cite{lub02, sho04, may02}), 
where intermediate nodes perform no 
additional coding and lower rates are achieved. 
It also differs from schemes that operate
in a block-by-block manner (e.g.,\  \cite{gdp04}),
where larger end-to-end delays are expected.
The main result of \cite{lme04} is that the scheme is indeed
capacity-achieving under the assumption of Poisson traffic with i.i.d.\
losses.

It turns out, however, that the scheme is in fact capacity-achieving
under much more general conditions, and one of our extensions is to show
that it can achieve capacity when the arrival of packets received on a
link is
described by any arrival process with an average rate.  No assumptions
are made on loss correlation or lack thereof---all we require is that a
long-run average rate of arrivals exists.  
This fact is particularly important in
wireless packet networks, where slow fading and collisions often cause
losses to be correlated in time and across separate links.
Our other main extension
is to derive error exponents in the case of Poisson traffic with i.i.d.\
losses.  Having error exponents is important as it allows us to quantify
the rate of decay of the probability of error with coding delay and to
determine the parameters of importance in this decay.

Our work draws upon results from the usually-disparate fields of
information theory and queueing theory and, to a degree, forms a
bridge between the two.  
In particular, we can view traditional error exponents as a limiting
case in which packets arrive in regular, deterministic intervals, 
and traditional queueing
as a special case in which no coding is performed.  Our approach relaxes
the constraints in these two limiting cases and can be looked upon as
a generalization to both.

\section{Network model}

We first describe a model for wireline packet networks.
We model a wireline packet network as a directed graph $\mathcal{G} =
(\mathcal{N},\mathcal{A})$,
where $\mathcal{N}$ is the set of nodes and $\mathcal{A}$ is the set of arcs.
Each arc $(i,j)$ represents a lossy point-to-point link.
Some subset of the packets injected into arc $(i,j)$ by node $i$ are
lost; the rest are received by node $j$ without error.
The process by which packets are injected into an arc is taken as given,
and it comes about, for example, by using some appropriate traffic
pattern (e.g.,\  deterministic or Poisson) at a rate that is
determined by optimizing 
according to some cost criterion 
over all rate allocations that allow for the
desired connection to be established 
(see \cite{lrm}).

We denote by $z_{ij}$ the average rate at which packets are
received on arc $(i,j)$.  More precisely, suppose that the arrival of
received packets on arc $(i,j)$ is described by the counting process
$A_{ij}$, i.e.\  for $\tau \ge 0$, $A_{ij}(\tau)$ is the total number of
packets received between time 0 and time $\tau$ on arc $(i,j)$.  
Then
$\lim_{\tau \rightarrow \infty} {A_{ij}(\tau)}/{\tau} = z_{ij}$ 
a.s.
We define a lossy wireline packet network as a pair $(\mathcal{G}, z)$.

We assume that links are delay-free in the sense that
the arrival time of a received packet corresponds
to the time that it was injected into the link.  We make this assumption
for the purpose of simplicity, and it can be relaxed.

For wireless packet networks, we model the network as a directed hypergraph 
$\mathcal{H} = (\mathcal{N},\mathcal{A})$, 
where $\mathcal{N}$ is the set of nodes and $\mathcal{A}$ is the set of
hyperarcs.  
A hypergraph is a generalization of a graph where generalized arcs,
called hyperarcs, connect two or more nodes.  Thus,
a hyperarc is a pair $(i,J)$, where $i$, the head, is an
element of $\mathcal{N}$, and $J$, the tail, is a non-empty subset of
$\mathcal{N}$.
Each hyperarc $(i,J)$ represents a lossy broadcast link.
For each $K \subset J$, some disjoint subset of the packets injected
into hyperarc $(i,J)$ by node $i$ are received by all nodes in $K$ without
error.

We denote by $z_{iJK}$ the average rate at which
packets, injected on hyperarc $(i,J)$, are received by the set of nodes $K
\subset J$.  More precisely, suppose that the arrival of packets that
are injected on hyperarc $(i,J)$ and received by the set of nodes $K
\subset J$ is described by the counting process $A_{iJK}$.  Then
$\lim_{\tau \rightarrow \infty} {A_{iJK}(\tau)}/{\tau} = z_{iJK}$ 
a.s.
We define a lossy wireless packet network as a pair $(\mathcal{H}, z)$.

\section{Coding scheme}
\label{sec:coding_scheme}

We suppose that, at the source node $s$, we have $K$ message packets
$w_1, w_2, \ldots, w_K$, which are vectors of length $\rho$ over the
finite field $\mathbb{F}_q$.  (If the packet length is $b$ bits, then we
take $\rho = \lceil b / \log_2 q \rceil$.)  The message packets are
initially present in the memory of node $s$.

The coding operation performed by each node is simple to describe and is
the same for every node:  Received packets are stored into the node's
memory, and packets are formed for injection with random
linear combinations of its memory contents 
whenever a packet injection occurs on an
outgoing hyperarc.  The coefficients of the combination are drawn uniformly
from $\mathbb{F}_q$.  

Since all coding is linear, we can write any
packet $x$ in the network as a linear combination of 
$w_1, w_2, \ldots, w_K$, namely,
$x = \sum_{k=1}^K \gamma_k w_k$. 
We call $\gamma$ the \emph{global
encoding vector} of $x$, and we assume that it is sent along with $x$,
in its header.  
The overhead this incurs (namely, $K \log_2 q$ bits)
is negligible if packets are sufficiently large.

A sink node collects packets and, if it has $K$ packets with
linearly-independent global encoding vectors, it is able to recover the
message packets.  Decoding can be done by Gaussian
elimination.  In addition, the scheme can be operated ratelessly, i.e.\  it
can be run indefinitely until all sink nodes in $T$ can decode (at which
stage that fact is signaled to all nodes, requiring only a small amount
of feedback).

\section{Coding theorems}

In this section, we specify achievable rate intervals for the coding
scheme in various scenarios.  
The fact that the intervals we specify are the largest possible
(i.e.\  that the scheme is capacity-achieving) follows from the 
cut-set bound for multi-terminal networks 
(see \cite[Section 14.10]{cot91}).

\subsection{Wireline networks}
\label{sec:coding_thms_wireline}

\subsubsection{Unicast connections}

Suppose that we wish to establish a connection of rate arbitrarily close
to $R$ packets per unit time from source node $s$ to sink node $t$.
Suppose further that
\[
R \leq \min_{Q \in \mathcal{Q}(s,t)}
  \left\{\sum_{(i, j) \in \Gamma_+(Q)} z_{ij}
  \right\},
\]
where $\mathcal{Q}(s,t)$ is the set of all cuts between $s$ and $t$, and
$\Gamma_+(Q)$ denotes the set of forward arcs of the cut $Q$, i.e.\  
\[
\Gamma_+(Q) := \{(i, j) \in \mathcal{A} \,|\, i \in Q, j \notin Q\} .
\]
Therefore, by the max-flow/min-cut theorem (see, for example,
\cite[Section 3.1]{ber98}), there exists a
flow vector $f$ satisfying
\[
\sum_{\{j | (i,j) \in \mathcal{A}\}} f_{ij} 
- \sum_{\{j | (j,i) \in \mathcal{A}\}} f_{ji} =
\begin{cases}
R & \text{if $i = s$}, \\
-R & \text{if $i = t$}, \\
0 & \text{otherwise},
\end{cases}
\]
for all $i \in \mathcal{N}$, and
$0 \le f_{ij} \le z_{ij}$
for all $(i,j) \in \mathcal{A}$.
We assume, without loss of generality, that $f$ is cycle-free in the
sense that the subgraph 
$\mathcal{G}^\prime = (\mathcal{N}, \mathcal{A}^\prime)$, where
$\mathcal{A}^\prime := \{(i,j) \in \mathcal{A} | f_{ij} > 0\}$, 
is acyclic.  (If $\mathcal{G}^\prime$
has a cycle, then it can be eliminated by subtracting flow from $f$ 
around it.)

Using the conformal realization theorem
(see, for example, \cite[Section 1.1]{ber98}), we decompose $f$ into
a finite set of paths $\{p_1, p_2, \ldots, p_M\}$, 
each carrying positive 
flow $R_{m}$ for $m= 1, 2, \ldots, M$, such that
$\sum_{m=1}^M R_{m} = R$.  

Our proof is based on tracking the propagation of what we call
\emph{globally innovative} packets.  
We first consider packets received by outward neighbors of $s$.  
Suppose packet $x$
is received by node $j$, an outward neighbor of $s$, at time
$\tau$.  We associate with $x$ the independent random variable $P_x$,
which, for $p_m$ containing arc $(s,j)$, takes the value
$p_m$ with probability $(1 - 1/q) R_m / z_{sj}$.
We then say that $x$ is globally innovative at node $j$ on path
$p_m$ if $P_x = p_m$.  
Suppose that the coding scheme is run for a total time $\Delta$, from
time 0 till time $\Delta$, and that, in this time, a total of $N$
globally innovative packets are received by outward neighbors of $s$.
We call these packets $v_1, v_2, \ldots, v_N$. 

In general,
any received packet $y$ in the network is a linear combination of packet
received by outward neighbors of $s$, so we can write
$y = \sum_{n=1}^N \beta_n v_n + \sum_{n=N+1}^{N^\prime} \delta_{n-N}v_n$,
where $v_{N+1}, v_{N+2}, \ldots, v_{N^\prime}$ denote the packets received by
outward neighbors of $s$ that are not globally innovative.  
Since $v_n$ is formed by a random linear combination of $w_1, w_2,
\ldots, w_K$, we have $v_n = \sum_{k=1}^K \alpha_{nk}w_k$ 
for $n = 1,2,\ldots, N$.  Hence, the $k$th component of the
global encoding vector of $y$ is given by
$\gamma_k = \sum_{n=1}^N \beta_n \alpha_{nk}
+ \sum_{n=N+1}^{N^\prime} \delta_{n-N} \alpha_{nk}$.
We call the vector $\beta$ associated with $y$ the \emph{auxiliary
encoding vector} of $y$, and we see that any sink that receives $K$ or
more packets with linear-independent auxiliary encoding vectors has $K$
packets whose global encoding vectors collectively form a random $K
\times K$ matrix over $\mathbb{F}_q$, with all entries chosen uniformly.
If this matrix is invertible, then the sink is able to recover the
message packets.  The probability that a random $K \times K$ matrix is
invertible is $\prod_{k=1}^K(1-1/q^k)$, which can be made
arbitrarily close to 1 by taking $q$ arbitrarily large.
Therefore, to determine whether a sink can recover the message packets,
we essentially need only to determine whether it receives $K$ or more
packets with linearly-independent auxiliary encoding vectors.

We have so far only defined the term globally innovative for packets
received by outward
neighbors of $s$.  Before defining the term for packets received by
the remaining nodes,
we associate with each path $p_m$ and each node $i$ on $p_m$ the
set of vectors $V_i^{(p_m)}$, which varies with time and
which is initially empty, i.e.\  $V_i^{(p_m)}(0) := \emptyset$.
Whenever a packet $x$ received at time $\tau$ is declared globally
innovative at node $i$ on path $p_m$, 
its auxiliary encoding vector $\beta$ is added to $V_i^{(p_m)}$,
i.e.\  $V_i^{(p_m)}(\tau^+) := \beta \cup V_i^{(p_m)}(\tau)$.
Let $U^{(p_m)} := V_j^{(p_m)}$, where $j$ is the outward neighbor of $s$
on $p_m$, and let $W^{(p_m)} := V_t^{(p_m)}$.
Now consider node $i \neq s$ and suppose packet $x$, with auxiliary
encoding vector $\beta$, is received by node $j$, an outward neighbor of
$i$, at time $\tau$.  
We again associate with $x$ the independent random
variable $P_x$, but $\Pr(P_x = p_m) = R_m / z_{ij}$ for $p_m$ containing arc
$(i,j)$, and we say that
$x$ is globally innovative at node $j$ on path $p_m$ if $P_x = p_m$, 
$\beta \notin \mathrm{span}(\cup_{l=1}^{m-1} W^{(p_l)}(\Delta)
\cup V_j^{(p_m)}(\tau) \cup \cup_{l=m+1}^M U^{(p_l)}(\Delta))$,
and $|V_i^{(p_m)}(\tau)| > |V_j^{(p_m)}(\tau)|$.

The definition of globally innovative is devised to satisfy two
conditions: first, that $\cup_{m=1}^M W^{(p_m)}(\Delta)$ is linearly
independent; and, second, that the propagation of globally innovative
packets through the network is described by a queueing network.
That the first of these two conditions is satisfied can be verified
easily:  Vectors are added to $W^{(p_1)}(\tau)$ only if they are
linearly independent of existing ones; vectors are added to
$W^{(p_2)}(\tau)$ only if they are linearly independent of existing ones
and ones in $W^{(p_1)}(\Delta)$; and so on.  The second of the two
conditions requires more investigation.

Consider path $p_m$ and node $i$ on $p_m$.  Let $j$ be the outward
neighbor of $i$ on $p_m$.  Suppose that 
packet $x$ is received by node $j$ from node $i$ at time $\tau$
and that
there are more globally innovative packets at $i$ than at $j$, 
i.e.\  $|V_i^{(p_m)}(\tau)| > |V_j^{(p_m)}(\tau)|$.   
Then, because 
$\cup_{l=1}^{m-1} W^{(p_l)}(\Delta) \cup V_i^{(p_m)}(\tau)
\cup \cup_{l=m+1}^M U^{(p_l)}(\Delta)$
and
$\cup_{l=1}^{m-1} W^{(p_l)}(\Delta) \cup V_j^{(p_m)}(\tau)
\cup \cup_{l=m+1}^M U^{(p_l)}(\Delta)$
are both linearly independent, 
$\mathrm{span}(\cup_{l=1}^{m-1} W^{(p_l)}(\Delta) \cup V_i^{(p_m)}(\tau)
\cup \cup_{l=m+1}^M U^{(p_l)}(\Delta)) 
\not\subset
\mathrm{span}(\cup_{l=1}^{m-1} W^{(p_l)}(\Delta) \cup V_j^{(p_m)}(\tau)
\cup \cup_{l=m+1}^M U^{(p_l)}(\Delta))$,
which implies that
$\mathrm{span}(V_i^{(p_m)}(\tau)) 
\not\subset
\mathrm{span}(\cup_{l=1}^{m-1} W^{(p_l)}(\Delta) \cup V_j^{(p_m)}(\tau)
\cup \cup_{l=m+1}^M U^{(p_l)}(\Delta))$.
Now, the packet $x$ is a random linear combination of vectors from a set
that contains $V_i^{(p_m)}(\tau)$, so $x$ is globally innovative with
some non-trivial probability.  This probability can be bounded using the
following lemma from \cite{dem}.  We quote the lemma without repeating
the proof.

\begin{Lem} 
\cite[Lemma 2.1]{dem}
Let $V_1$ and $V_2$ be two collections of vectors from
$\mathbb{F}_q^n$, and let $\beta$ be a random linear combination of the
vectors in $V_1$, with the coefficients of the combination drawn
uniformly from $\mathbb{F}_q$.  Then
\[
\Pr(\beta \notin \mathrm{span}(V_2) \,|\, \mathrm{span}(V_1)
\not\subset \mathrm{span}(V_2)) \ge 1 - \frac{1}{q} .
\]
\label{lem:100}
\end{Lem}

It follows from Lemma~\ref{lem:100} that $x$ is globally innovative with
probability not less than $(1 - 1/q)R_m / z_{ij}$.
Since we can always discard globally innovative packets, we assume that
$x$ is globally innovative with probability exactly
$(1-1/q)R_m / z_{ij}$.
If instead $|V_i^{(p_m)}(\tau)| = |V_j^{(p_m)}(\tau)|$, we see that
$x$ cannot be globally innovative, and this remains true until another
arrival occurs at $i$.  Therefore, the propagation of innovative packets
through node $i$ on path $p_m$ can be described as the propagation of
jobs through a single-server queueing station, where
the state $|V_i^{(p_m)}(\tau)| > |V_j^{(p_m)}(\tau)|$ corresponds to a
non-empty queue and 
the state $|V_i^{(p_m)}(\tau)| = |V_j^{(p_m)}(\tau)|$ corresponds to an
empty queue.

The queueing station is serviced with probability $(1 - 1/q)R_m /z_{ij}$
whenever the queue is non-empty and a received packet arrives on arc
$(i,j)$.  We can equivalently consider ``candidate'' packets that arrive
with probability $(1-1/q) R_m/z_{ij}$ whenever a received packet arrives
on arc $(i,j)$ and say that the queueing station is serviced whenever the
queue is non-empty and a candidate packet arrives on arc $(i,j)$.
Therefore, the queueing network that we wish to analyze is one with $N$
jobs initially present at node $s$ and with $M$ paths where, on each
path, jobs are serviced on each arc by the arrival of candidate packets.

We analyze the queueing network of interest using the fluid
approximation for discrete-flow networks (see, for example, \cite{chy01,
chm91}).
We begin by considering a single path $p_m$.
We write $p_m = \{i_1, i_2, \ldots, i_{L_m}, t\}$,
where $i_1 = s$.
Let $B_{ml}$ and $C_{ml}$ be the counting processes for the arrival of 
globally innovative packets and candidate packets, respectively,
on arc $(i_l, i_{l+1})$ for path $p_m$.
Let $Q_{ml}^{(N_m)}(\tau)$ be the number of packets queued for service 
at $i_l$ at time $\tau$ when there are $N_m$ jobs initially present at
node $s$.
Hence, for $l = 1, 2, \ldots, L_m$,
$Q_{ml}^{(N_m)} = B_{m(l-1)} - B_{ml}$,
where $B_{m0}(\tau) := N_m$ for all $\tau \ge 0$.
Let $C_{m0}(\tau) := N_m$ for all $\tau \ge 0$,
$X_{ml} := C_{m(l-1)} - C_{ml}$, and $Y_{ml} := C_{ml} - B_{ml}$.  Then,
for $l = 1, 2, \ldots, L_m$,
\begin{equation}
Q_{ml}^{(N_m)} 
= X_{ml} - Y_{m(l-1)} + Y_{ml} .
\label{eqn:200}
\end{equation}
Moreover, we have
\begin{gather}
Q_{ml}^{(N_m)}(\tau) dY_{ml}(\tau) = 0, \\
dY_{ml}(\tau) \ge 0,
\end{gather}
and
\begin{equation}
Q_{ml}^{(N_m)}(\tau) \ge 0
\end{equation}
for all $\tau \ge 0$ and $l = 1, 2, \ldots, L_m$, and
\begin{equation}
Y_{ml}(\tau) = 0
\label{eqn:210}
\end{equation}
for all $l = 1, 2, \ldots, L_m$.

We observe now that 
equations (\ref{eqn:200})--(\ref{eqn:210}) give us
the conditions for a Skorohod problem (see,
for example, \cite[Section 7.2]{chy01}) and, by the oblique reflection
mapping theorem, there is a well-defined, 
Lipschitz-continuous mapping $\Phi$ such that $Q^{(N_m)}_m = \Phi(X_m)$.  

Let 
$\bar{C}^{(N_m)}_{ml}(\tau) := C_{ml}(N_m \tau)/N_m$, 
$\bar{X}^{(N_m)}_{ml}(\tau) := X_{ml}(N_m \tau)/N_m$, 
and
$\bar{Q}^{(N_m)}_{ml}(\tau) := Q_{ml}^{(N_m)}(N_m \tau)/N_m$.
Recall that $A_{ij}$ is the counting process
for the arrival of received packets on arc $(i,j)$.  Therefore,
$C_{ml}^{(N_m)}(\tau)$ is the sum of $A_{i_li_{l+1}}(\tau)$ 
Bernoulli-distributed random
variables with parameter $(1-1/q)R_m/z_{i_li_{l+1}}$.  
Hence
\[
\begin{split}
\bar{C}_{ml}(\tau) 
&:= 
\lim_{N_m \rightarrow \infty}
\bar{C}^{(N_m)}_{ml}(\tau) \\
& = 
\lim_{N_m \rightarrow \infty}
\frac{1}{N_m} \frac{(1-1/q)R_m}{z_{i_li_{l+1}}}
A_{i_li_{l+1}}(N_m \tau) 
\qquad \text{a.s.} \\
&=
(1-1/q)R_m \tau
\qquad \text{a.s.},
\end{split}
\]
where the last equality follows by the assumptions of the model.
Therefore,
\[
\begin{split}
\bar{X}_{ml}(\tau) &:=
\lim_{N_m \rightarrow \infty}
\bar{X}_{ml}^{(N_m)}(\tau) \\
&=
\begin{cases}
1 - (1-1/q)R_m \tau & \text{if $l = 1$}, \\
0 & \text{otherwise} 
\end{cases}
\qquad \text{a.s.}
\end{split}
\]
By the Lipschitz-continuity of $\Phi$, then, it follows that
$\bar{Q}_m := \lim_{N_m \rightarrow \infty} \bar{Q}^{(N_m)}_m 
= \Phi(\bar{X}_m)$,  i.e.\
$\bar{Q}_m$ is, almost surely, 
the unique $\bar{Q}_m$ that satisfies, for some
$\bar{Y}_m$,
\begin{gather}
\bar{Q}_{ml}(\tau) =
\begin{cases}
1 - (1-1/q)R_m \tau + \bar{Y}_{m1}(\tau) & \text{if $l = 1$}, \\
\bar{Y}_{ml}(\tau) - \bar{Y}_{m(l-1)}(\tau) & \text{otherwise} ,
\end{cases} 
\label{eqn:300} \\
\bar{Q}_{ml}(\tau) d\bar{Y}_{ml}(\tau) = 0, \\
d\bar{Y}_{ml}(\tau) \ge 0, 
\end{gather}
and
\begin{equation}
\bar{Q}_{ml}(\tau) \ge 0
\end{equation}
for all $\tau \ge 0$ and $l = 1,2, \ldots, L_m$, and
\begin{equation}
\bar{Y}_{ml}(0) = 0
\label{eqn:310}
\end{equation}
for all $l = 1,2,\ldots, L_m$.

A pair $(\bar{Q}_m, \bar{Y}_m)$ that satisfies
(\ref{eqn:300})--(\ref{eqn:310}) is 
\begin{equation}
\bar{Q}_{ml}(\tau)
= 
\begin{cases}
\left(
1 - (1-1/q)R_m \tau
\right)^+
& \text{if $l=1$}, \\
0 & \text{otherwise},
\end{cases}
\label{eqn:320}
\end{equation}
and
\[
\bar{Y}_{ml}(\tau)
= 
\left(
1 - (1-1/q)R_m \tau
\right)^- .
\]
Hence $\bar{Q}_m$ is given by equation (\ref{eqn:320}).

Recall that the sink can recover the message packets with high
probability if it receives $K$ or more packets with linearly-independent
auxiliary encoding vectors and that $\cup_{m=1}^M W^{(p_m)}(\Delta)$ is
linearly-independent.  Therefore, the sink can recover the message
packets with high probability if it has $K$ or more globally innovative
packets at $t$ on any path $p_m$.  
Let $\nu$ be the number of 
globally innovative packets at $t$ at time $\Delta$.  Then
$\nu \ge N - \sum_{m=1}^M \sum_{l=1}^{L_m} Q^{(N_m)}_{ml}(\Delta)$,
where $N_m = C_{m1}(\Delta)$.
Take $\Delta = N/(1-1/q)R$.
Then
\[
\begin{split}
\lim_{N \rightarrow \infty} \frac{\nu}{N}
&\ge \lim_{N \rightarrow \infty} 1 - 
\sum_{m=1}^M \sum_{l=1}^{L_m} \frac{N_m}{N} 
\bar{Q}^{(N_m)}_{ml}\left(\frac{N}{N_m(1-1/q)R}\right) \\
&= 1
\qquad \text{a.s.},
\end{split}
\]
since $\lim_{N \rightarrow \infty} N_m/N = R_m/R$ a.s.

Take $K = \lceil N R_c \rceil$, where $0 \le R_c < 1$.
Then, for $N$ sufficiently large,
$\nu \ge K$ with probability arbitrarily close to 1.
So the probability of error can be made arbitrarily small, and the rate
achieved is
\[
\frac{K}{\Delta} = 
\frac{K}{N} (1-1/q)R
\ge
R_c(1-1/q) R ,
\]
which can be made arbitrarily close to $R$.
This concludes the proof for unicast connections.

\subsubsection{Multicast connections}

The proof for multicast connections is very similar to that for unicast
connections.  In this
case, rather than a single sink $t$, we have a set of sinks $T$.  And
we suppose
\[
R \leq \min_{t \in T}\min_{Q \in \mathcal{Q}(s,t)}
  \left\{\sum_{(i, j) \in \Gamma_+(Q)} z_{ij}
  \right\}.
\]
Therefore, by the max-flow/min-cut theorem, there exists, for all 
$t \in T$, a flow vector $f^{(t)}$ satisfying
\[
\sum_{\{j | (i,j) \in \mathcal{A}^\prime\}} f_{ij}^{(t)} 
- \sum_{\{j | (j,i) \in \mathcal{A}^\prime\}} f_{ji}^{(t)} =
\begin{cases}
R & \text{if $i = s$}, \\
-R & \text{if $i = t$}, \\
0 & \text{otherwise},
\end{cases}
\]
for all $i \in \mathcal{N}$, and
$f_{ij}^{(t)} \le z_{ij}$ for all $(i,j) \in \mathcal{A}^\prime$.  

For each flow vector $f^{(t)}$, we go through the same argument as that
for a unicast connection, and we find that the probability of error at
every sink node can be made arbitrarily small by taking $N$ and $q$
sufficiently large.

We summarize our results regarding wireline networks with the following
theorem statement.

\begin{Thm}
Consider the lossy wireline packet network $(\mathcal{G}, z)$.
The random linear coding scheme we describe is capacity-achieving for
multicast connections,
i.e.\  it can achieve, with
arbitrarily small error probability, a multicast
connection over the network
from source node $s$ to sink nodes in the set $T$ at rate
arbitrarily close to $R$ packets per unit time if
\[
R \leq \min_{t \in T} \min_{Q \in \mathcal{Q}(s,t)}
  \left\{\sum_{(i, j) \in \Gamma_+(Q)} z_{ij}
  \right\}.
\]
\label{thm:100}
\end{Thm}

\subsection{Wireless packet networks}

The wireless case is actually very similar to the wireline one.
The main difference is that
the same packet may be received by more than one
node, so we must specify which node considers the packet to be
globally innovative.  
With an appropriate redefinition of globally innovative, we
obtain the following theorem.
We omit the details of the development for the sake of brevity.

\begin{Thm}
Consider the lossy wireless packet network $(\mathcal{H}, z)$.
The random linear coding scheme we describe is capacity-achieving for
multicast connections,
i.e.\  it can achieve, with
arbitrarily small error probability, a multicast
connection over the network
from source node $s$ to sink nodes in the set $T$ at rate
arbitrarily close to $R$ packets per unit time if
\[
R \leq \min_{t \in T} \min_{Q \in \mathcal{Q}(s,t)}
  \left\{\sum_{(i, J) \in \Gamma_+(Q)} \sum_{K \not\subset Q} z_{iJK}
  \right\}.
\]
\end{Thm}

\section{Error exponents for Poisson traffic with i.i.d.\  losses}

We now look at the rate of decay of the probability of
error $p_e$ in the coding delay $\Delta$.  
In contrast to traditional error exponents where coding delay is
measured in symbols, we measure coding delay in time units---time
$\tau = \Delta$ is 
the time at which the sink nodes attempt to decode the
message packets.  The two methods of measuring delay are essentially
equivalent when packets arrive in regular, deterministic intervals.

We specialize to the case of Poisson traffic with i.i.d.\  losses.  
Hence, in the wireline case, the process $A_{ij}$ is a Poisson process
with rate $z_{ij}$ and, in the wireless case, the process $A_{iJK}$ is a
Poisson process with rate $z_{iJK}$.

Consider the unicast case for now, and
suppose we wish to establish a connection of rate $R$.
Let $C$ be the supremum of all asymptotically-achievable rates.
We consider the limit where $q$, the coding
field size, is infinitely large, thus implying large coding
complexity and packet size.     

We begin by deriving an upper bound on the probability of error.
To this end, we take a flow vector $f$ from $s$ to $t$ of size $C$
and develop a queueing network from it that describes the propagation of
globally innovative packets.
This queueing network now becomes a Jackson network.
Suppose that $N$, the number of jobs initially present at $s$, 
is infinitely large.  Then, as a consequence of Burke's
theorem (see, for example, \cite[Section 2.1]{kel79}) and the fact that
the queueing network is acyclic, the
arrival and departure processes at all nodes are 
Poisson in steady-state.  

Let $\Psi_{t}(m)$ be the arrival time of the $m$th globally innovative packet
at $t$.  Then, we have, for $\theta < C$,
\begin{equation}
\lim_{m \rightarrow \infty} \frac{1}{m}
\log \mathbb{E}[\exp(\theta \Psi_{t}(m))]
= \log \frac{C}{C - \theta} ,
\label{eqn:100}
\end{equation}
since, in steady-state, 
the arrival of globally innovative packets at $t$ is described by a Poisson
process of rate $C$.
If an error occurs, then fewer than $\lceil R\Delta \rceil$
globally innovative packets are received by $t$ by 
time $\tau = \Delta$, which is
equivalent to
saying that $\Psi_{t}(\lceil R\Delta \rceil) > \Delta$.
Therefore,
\[
p_e \le \Pr(\Psi_{t}(\lceil R\Delta \rceil) > \Delta),
\]
and, using the Chernoff bound, we obtain
\[
p_e \le \min_{0 \le \theta < C}
\exp\left(
-\theta \Delta + \log \mathbb{E}[\exp(\theta \Psi_{t}(\lceil R\Delta
\rceil) )] 
\right) .
\]
Let $\varepsilon$ be a positive real number.  
Then using equation (\ref{eqn:100}) we obtain, 
for $\Delta$ sufficiently large,
\[
\begin{split}
p_e &\le \min_{0 \le \theta < C}
\exp\left(-\theta \Delta
+ R \Delta \left\{\log \frac{C}{C-\theta} + \varepsilon \right\} \right)
\\
&= \exp( -\Delta(C-R-R\log(C/R)) + R\Delta \varepsilon) .
\end{split}
\]
Hence, we conclude that
\begin{equation}
\lim_{\Delta \rightarrow \infty} \frac{-\log p_e}{\Delta}
\ge C - R - R\log(C/R) .
\label{eqn:110}
\end{equation}

For the lower bound, we examine 
a cut whose flow capacity is $C$.  We take one such cut and denote it by
$Q^*$.  It is
clear that, if fewer than $\lceil R\Delta \rceil$ distinct packets are
received across $Q^*$ in time $\tau = \Delta$, then an error occurs.
For both wireline and wireless networks, the arrival of
distinct packets across $Q^*$ is described by a Poisson
process of rate $C$.  
Thus we have
\[
\begin{split}
p_e &\ge \exp(-C\Delta)
\sum_{l = 0}^{\lceil R\Delta \rceil - 1}
\frac{(C\Delta)^l}{l!}  \\
&\ge \exp(-C \Delta)
\frac{(C\Delta)^{\lceil R\Delta \rceil -1}}
{\Gamma(\lceil R \Delta \rceil)} ,
\end{split}
\]
and, using Stirling's formula, we obtain the reverse inequality to
(\ref{eqn:110}).  
Therefore, 
\begin{equation}
\lim_{\Delta \rightarrow \infty} \frac{-\log p_e}{\Delta}
= C - R - R\log(C/R) .
\label{eqn:120}
\end{equation}

Equation (\ref{eqn:120}) defines the asymptotic rate of decay of the
probability of error in the coding delay $\Delta$.
This asymptotic rate of decay is
determined entirely by $R$ and $C$.  Thus, for a packet network with
Poisson traffic employing the coding scheme described in
Section~\ref{sec:coding_scheme} with large field size $q$, the flow
capacity $C$ of the minimum cut of the network is essentially
the sole figure of merit of importance in determining the effectiveness
of the coding scheme for large, but finite, coding delay.
Thus, in deciding how to inject
packets to support the desired connection, 
we need only consider this figure of merit.

Extending the result from unicast connections to multicast connections
is quite straightforward.  We again obtain equation (\ref{eqn:120}) in
the multicast case.

\bibliographystyle{IEEEtran}
\bibliography{IEEEabrv,inform_theory}

\begin{thebibliography}{10}
\providecommand{\url}[1]{#1}
\csname url@rmstyle\endcsname
\providecommand{\newblock}{\relax}
\providecommand{\bibinfo}[2]{#2}
\providecommand\BIBentrySTDinterwordspacing{\spaceskip=0pt\relax}
\providecommand\BIBentryALTinterwordstretchfactor{4}
\providecommand\BIBentryALTinterwordspacing{\spaceskip=\fontdimen2\font plus
\BIBentryALTinterwordstretchfactor\fontdimen3\font minus
  \fontdimen4\font\relax}
\providecommand\BIBforeignlanguage[2]{{%
\expandafter\ifx\csname l@#1\endcsname\relax
\typeout{** WARNING: IEEEtran.bst: No hyphenation pattern has been}%
\typeout{** loaded for the language `#1'. Using the pattern for}%
\typeout{** the default language instead.}%
\else
\language=\csname l@#1\endcsname
\fi
#2}}

\bibitem{lme04}
D.~S. Lun, M.~M{\'e}dard, and M.~Effros, ``On coding for reliable communication
  over packet networks,'' in \emph{Proc. 42nd Annual Allerton Conference on
  Communication, Control, and Computing}, Sept.--Oct. 2004, invited paper.

\bibitem{lub02}
M.~Luby, ``{LT} codes,'' in \emph{Proc. 43rd Annual IEEE Symposium on
  Foundations of Computer Science}, Nov. 2002, pp. 271--280.

\bibitem{sho04}
A.~Shokrollahi, ``Raptor codes,'' Jan. 2004, preprint.

\bibitem{may02}
P.~Maymounkov, ``Online codes,'' NYU, Technical Report TR2002-833, Nov. 2002.

\bibitem{gdp04}
R.~Gowaikar, A.~F. Dana, R.~Palanki, B.~Hassibi, and M.~Effros, ``On the
  capacity of wireless erasure networks,'' in \emph{Proc. 2004 IEEE
  International Symposium on Information Theory (ISIT 2004)}, Chicago, IL,
  June--July 2004, p. 401.

\bibitem{lrm}
D.~S. Lun, N.~Ratnakar, M.~M{\'e}dard, R.~Koetter, D.~R. Karger, T.~Ho, and
  E.~Ahmed, ``Minimum-cost multicast over coded packet networks,'' submitted to
  \emph{IEEE Trans. Inform. Theory}.

\bibitem{cot91}
T.~M. Cover and J.~A. Thomas, \emph{Elements of Information Theory}.\hskip 1em
  plus 0.5em minus 0.4em\relax New York, NY: John Wiley \& Sons, 1991.

\bibitem{ber98}
D.~P. Bertsekas, \emph{Network Optimization: Continuous and Discrete
  Models}.\hskip 1em plus 0.5em minus 0.4em\relax Belmont, MA: Athena
  Scientific, 1998.

\bibitem{dem}
S.~Deb and M.~M\'edard, ``Algebraic gossip: A network coding approach to
  optimal multiple rumor mongering,'' submitted to \emph{IEEE Trans. Inform.
  Theory}.

\bibitem{chy01}
H.~Chen and D.~D. Yao, \emph{Fundamentals of Queueing Networks: Performance,
  Asymptotics, and Optimization}, ser. Applications of Mathematics.\hskip 1em
  plus 0.5em minus 0.4em\relax New York, NY: Springer, 2001, vol.~46.

\bibitem{chm91}
H.~Chen and A.~Mandelbaum, ``Discrete flow networks: Bottleneck analysis and
  fluid approximations,'' \emph{Math. Oper. Res}, vol.~16, no.~2, pp. 408--446,
  May 1991.

\bibitem{kel79}
F.~P. Kelly, \emph{Reversibility and Stochastic Networks}.\hskip 1em plus 0.5em
  minus 0.4em\relax Chichester: John Wiley \& Sons, 1979.

\end{thebibliography}

\end{document}